\documentclass[amsmath,amssymb,twocolumn,groupedaddress]{revtex4-1}
\usepackage{graphicx}
\usepackage{pstricks}
\usepackage{epsf}
\usepackage{natbib}
\newcommand{\be}{\begin{equation}}\newcommand{\ee}{\end{equation}}
\newcommand{\ba}{\begin{array}{l}}\newcommand{\ea}{\end{array}}
\newcommand{\baa}{\begin{eqnarray}}\newcommand{\eaa}{\end{eqnarray}}
\newcommand{\lab}[1]{\label{#1}}\newcommand{\re}[1]{(\ref{#1})}
\newcommand{\ci}[1]{\cite{#1}}

\def\pa{\partial}

\begin{document}
\title{Sine-Gordon solitons in networks: Scattering and transmission at
vertices}
\author{Zarif Sobirov$^{a,c}$\thanks{sobirovzar@gmail.com}, Doniyor Babajanov$^b$, Davron Matrasulov$^b$, Katsuhiro Nakamura$^{d,e}$, Hannes Uecker$^f$\thanks{hannes.uecker@uni-oldenburg.de}}

\affiliation{
$^a$ Tashkent Financial Institute, 60A, Amir Temur Str., 100000, Tashkent, Uzbekistan\\
$^b$ Turin Polytechnic University in Tashkent, 17 Niyazov Str.,
100095,  Tashkent, Uzbekistan\\
$^c$Faculty of Mathematics, National University of Uzbekistan, Vuzgorodok, Tashkent 100174,Uzbekistan\\
$^d$Faculty of Physics, National University of Uzbekistan, Vuzgorodok, Tashkent 100174,Uzbekistan\\
$^e$Department of Applied Physics, Osaka City University, Osaka 558-8585, Japan\\
$^f$ Institut f\"ur Mathematik, Universit\"at Oldenburg, D26111
Oldenburg, Germany}

\begin{abstract}
  We consider the sine-Gordon equation on metric graphs  with
  simple topologies and derive vertex boundary conditions
 from the fundamental conservation laws together with successive space-derivatives of sine-Gordon equation.  We analytically
obtain traveling wave solutions  in the form of standard sine-Gordon
solitons such as kinks and antikinks for star and
tree graphs. We show that for this case the sine-Gordon equation becomes completely integrable just as in case of a simple $1D$ chain.
This simple analysis provides a cornerstone for
the numerical solution of the general case, including a
quantification of  the vertex scattering.
Applications of  the obtained results to Josephson junction networks and DNA double helix are discussed.
\end{abstract}
\pacs{05.45.Yv,02.30.Ik,42.65.Tg}
\maketitle

\vskip 0.2cm
\textit{Introduction.}
Nonlinear wave dynamics described by the sine-Gordon equation is
of importance in a variety of topics in physics, such as as elastic and stress
wave propagation in solids, liquids and tectonic plates (see,
e.g., \ci{ablowitz1,Raj, drazin, ablowitz,Kivshar,
dauxois, Panos}), transport in Josephson junctions
\ci{Baron}, and topological quantum fields \ci{Raj,dauxois}.
Continuous and discrete forms of the sine-Gordon equation have been
used so far for the description of wave transport in different
media. However, there are structures for which the wave dynamics
cannot be described within the
traditional continuous or discrete approaches. These are networks and branched
structures where the transmission through a branching point
(network vertex) should be described by vertex conditions. Early
studies of nonlinear evolution equations in branched structures are
\ci{New1,New2,New3}, and in
recent few years one can observe rapidly growing interest in
nonlinear waves and soliton transport in networks described by
nonlinear Schr\"odinger equation
\ci{zar2010,adami2011,adami-eur,adami-jpa,adami2013,Karim2013}.
Integrable boundary conditions following from the
conservation laws were formulated, and soliton solutions
 yielding reflectionless transport across the graph vertex
were derived in \ci{zar2010},  see also \ci{Zarif2011} for the
case of a  discrete nonlinear Schr\"odinger equation.
Burioni {\it et al} \ci{Burioni1,Burioni2}
studied the discrete nonlinear Schr\"odinger equation and computed
transport and reflection coefficients as a function of the
wavenumber of a Gaussian wave packet and the length of a graph
attached to a defect site.

In this paper we address the  wave dynamics in networks
described by the sine-Gordon equation on metric graphs, which can
be used for modeling of soliton transport in DNA
double helix, tectonic plates and Josephson junction networks. The latter has
attracted much attention in condensed matter physics
\ci{sodano-np09,sodano13}. Another interesting application of sine-Gordon equations,
 or, more generally, nonlinear Klein-Gordon equations,
on metric graphs can be networks of granular chains
\ci{New4,New3}.
Recently, soliton dynamics in networks was studied
by considering the 2D sine-Gordon equation on $Y$ and $T$
junctions \ci{caputo14}, and the metric graph limit was also
studied numerically. See also \cite{UGSBM15} for similar results
for the 2D Nonlinear Schr\"odinger equation on ``fat'' graphs.

Here we focus on the problem of integrability of sine-Gordon equations
on metric graphs and soliton transmission at the graph vertex. In
particular, using an approach similar to that of \ci{zar2010},  we  discuss
the conditions under which the
sine-Gordon equation is completely  integrable and allows exact traveling wave solutions which provide reflectionless
transmission of sine-Gordon solitons across vertices. Numerical
solutions with scattering at a vertex when these conditions are violated
are also presented.

\textit{Conservation laws and boundary conditions.}
For evolution equations on graphs, the connections of
the bonds at the vertices are provided by vertex boundary conditions.
In case of linear wave equations such conditions follow from
self-adjointness of the problem \ci{Kostrykin99, EK15}.
For nonlinear evolution equations one should use such fundamental laws as energy, flux, momentum and (for sine-Gordon model) topological charge conservations \ci{zar2010,adami2011,caputo14}. Below we derive such conditions and show the existence of infinitely many conservation laws in our model, which yields the complete integrability of the system.

Most of the 1D sine-Gordon models follow from the Lagrangian density
$ L=\left[
  \frac{1}{2}\left(u_{t}^2-a^2u_{x}^2\right)-\beta(1-\cos u)\right],
$
where  $a$ and $\beta$ are positive constants.
\begin{figure}[ht!]
 \includegraphics[width=90mm]{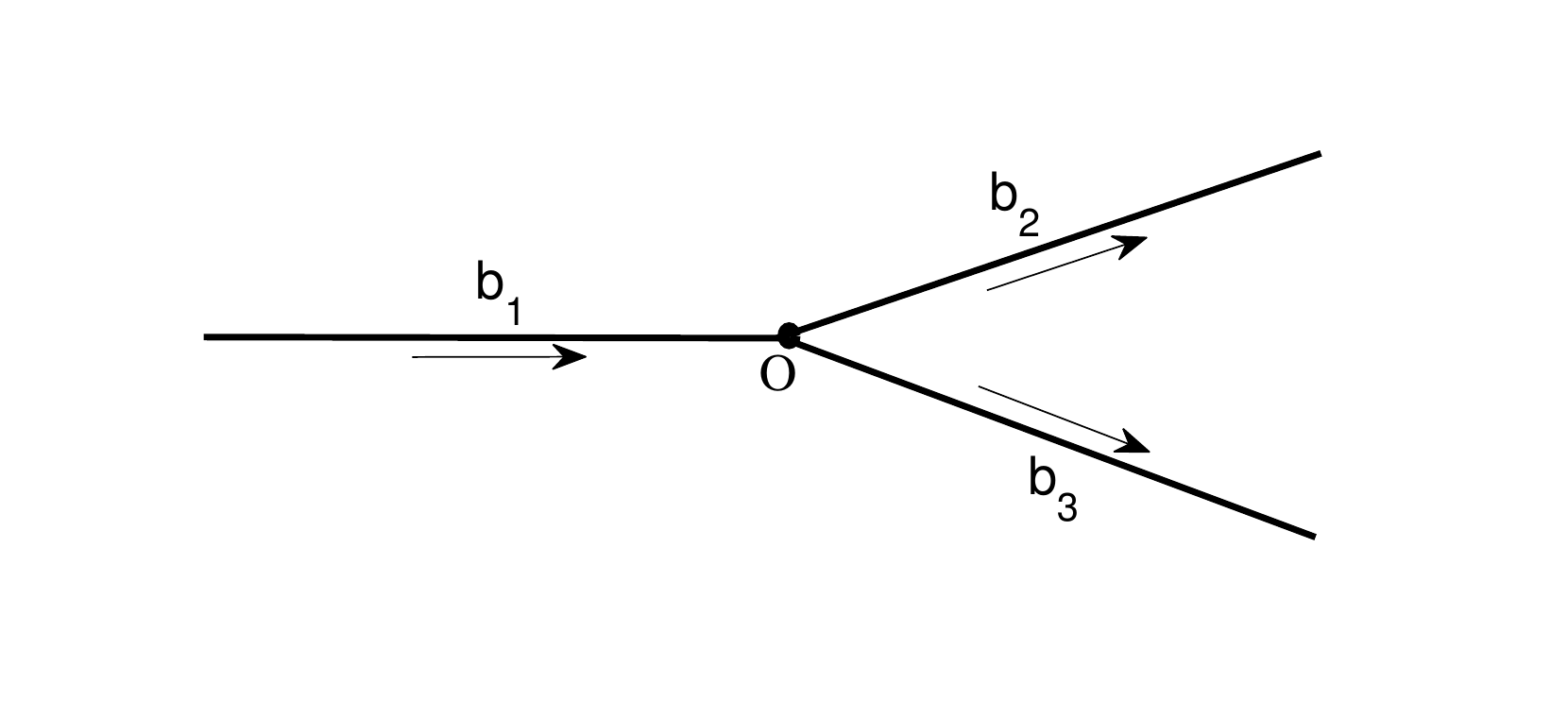}\\[-10mm]
\caption{Sketch of a metric  star graph \label{pic1}}
\end{figure}
We
want to explore a sine-Gordon model on networks modeled by
graphs, i.e., system of bonds which are connected at one
or more vertices (branching points). The connection rule is called the
topology of the graph. When the bonds can be
assigned a length, the graph is called a metric
graph. The sine-Gordon model on each bond $b_k$,
$k=1,2,3,..., N$ is given by Lagrangian density
$$
L_k = \left[
  \frac{1}{2}\left(u_{kt}^2-a_k^2u_{kx}^2\right)-\beta_k
(1-\cos u_k)\right],
$$
where 
$a_k, \beta_k >0$.  In the following we consider
a star graph with three semi-infinite bonds connected at the
point $O$ called vertex of the graph, see Fig.~1. The
coordinates are defined as $x_1\in(-\infty,0]$ and
$x_{2,3}\in[0, \infty)$, where 0 corresponds to
the vertex point. The equation of motion derived from the
above Lagrangian density leads to the sine-Gordon equation on each bond given as
\begin{equation}\label{eq1}
  u_{ktt}-a_k^2u_{kxx}+\beta_k\sin u_k=0.
\end{equation}
To formulate physically reasonable vertex boundary conditions (VBC) one can use
the continuity of wave function
\begin{equation}\label{r-cont}
  u_1(0,t)=u_2(0,t)=u_3(0,t)
\end{equation}
and fundamental conservation laws such as energy, charge and momentum conservations together with the asymptotic conditions at infinities:
$\pa_x u_1(x_1,t), \pa_t u_1(x_1,t) {\to}0$ and
$u_1(x_1,t)\to 2\pi n_1$ as $x_1\to-\infty$, and
$\pa_x u_k(x_k,t), \pa_t u_k(x_k,t) \to 0$ and
$u_k(x_k,t)\to 2\pi n_k$ as  $x_{k}\to\infty, k=2,3$, for some integer $n_k$, $k=1,2,3$.

For the primary star graph in Fig \ref{pic1}, the energy is defined as 
\begin{equation}\label{energy}
E(t)=\sum_{k=1}^{3}\frac{1}{\beta_k}\int\limits_{B_k}\left[
\frac{1}{2}\left(u_{kt}^2+a_k^2u_{kx}^2\right)+\beta_k(1-\cos u_k)\right]dx,
\end{equation}
where  $B_1{=}(-\infty,0), \ B_{2,3}{=}(0,+\infty)$.
Then
\begin{equation*}
\dot{E}=\left.\frac{a_1^2}{\beta_1}u_{1x}u_{1t}\right|_{x_1=0}-\left.\frac{a_2^2}{\beta_2}u_{2x}u_{2t}\right|_{x_2=0}-
  \left.\frac{a_3^2}{\beta_3}u_{3x}u_{3t}\right|_{x_3=0},
\end{equation*}
and by (\ref{r-cont}) the energy conservation reduces to
\begin{equation}\label{VC-E}
\left.\frac{a_1^2}{\beta_1}u_{1x}\right|_{x_1=0}=\left.\frac{a_2^2}{\beta_2}u_{2x}\right|_{x_2=0}+
\left.\frac{a_3^2}{\beta_3}u_{3x}\right|_{x_3=0}.
\end{equation}

For the same star graph as in Fig \ref{pic1}, the charge $Q$ is given by
\begin{equation}\label{topol-index}
  2\pi Q=\frac{a_1}{\sqrt{\beta_1}}\int\limits_{-\infty}^{0}u_{1x}dx+\sum_{k=2}^{3}\frac{a_k}{\sqrt{\beta_k}}\int\limits_{0}^{+\infty}u_{kx}dx.
\end{equation}
From $\dot{Q}=0$ and (\ref{r-cont}) we obtain the sum rule
\begin{equation}\label{r-sum0}
\frac{a_1}{\sqrt{\beta_1}}=\frac{a_2}{\sqrt{\beta_2}}+\frac{a_3}{\sqrt{\beta_3}}.
\end{equation}

The initial boundary problem (IBVP) (\ref{eq1}), (\ref{r-cont})
and (\ref{VC-E}) with appropriate initial conditions and
asymptotic conditions at infinities is now well defined. However,
to see the infinite number of constants of motion, we must search
for other additional conditions for parameters.

\textit{Integrability and traveling wave solutions.}
We consider the momentum defined by
\begin{equation}\label{r-momen}
P=\sum_{k=1}^3 \frac{a_k}{\beta_k}\int_{B_k}u_{kx}u_{kt}dx,
\end{equation}
such that
\begin{eqnarray}\label{r-momen2}
\dot{P}&=&\frac{a_1}{\beta_1}\left.\left[\frac{1}{2}(u_{1t}^2+a_1^2u_{1x}^2)-\beta_1(1{-}\cos u_1)\right]\right|_{x_1=0}\nonumber\\
&-&
\sum_{k=2}^3 \frac{a_k}{\beta_k}\left.\left[\frac{1}{2}(u_{kt}^2{+}a_k^2u_{kx}^2)
{-}\beta_k(1{-}\cos u_k)\right]\right|_{x_k=0}.
\end{eqnarray}
For $\dot{P}=0$, we impose  the condition,
\begin{equation}\label{r-momen3}
\beta_1=\beta_2=\beta_3=\beta(>0),
\end{equation}
which simplifies the sum rule (\ref{r-sum0}) to
\begin{equation}\label{r-sum}
a_1=a_2+a_3.
\end{equation}
Then (\ref{r-momen2}) becomes
\begin{eqnarray*}
2\beta\dot{P}&=&a_1^3u_{1x}^2(0,t)- a_2^3u_{2x}^2(0,t) -a_3^3u_{3x}^2(0,t)\\
&=& -\frac{a_2a_3}{a_2+a_3}(a_2u_{2x}(0,t)-a_3u_{3x}(0,t))^2,
\end{eqnarray*}
where we used (\ref{VC-E}), (\ref{r-momen3}) and (\ref{r-sum}) in obtaining
the last expression.
Thus, $\dot P=0$ yields
$a_2u_{2x}(0,t)=a_3u_{3x}(0,t)$, which together with
(\ref{VC-E}), (\ref{r-momen3}) and (\ref{r-sum}) gives
\begin{equation}\label{r-deriv1}
a_1u_{1x}(0,t)=a_2u_{2x}(0,t)=a_3u_{3x}(0,t).
\end{equation}

Conditions on higher-order space-derivatives may be available from
higher-order conservations, where the analysis becomes more laborious.
However, they can also be obtained directly from (\ref{eq1}), (\ref{r-cont}) via
$\left.a_1^2u_{1xx}\right|_{x_1=0}=\left.(u_{1tt}+\beta\sin u_1)\right|_{x_1=0}
=\left.(u_{ktt}+\beta\sin u_k)\right|_{x_k=0}=\left.a_k^2u_{kxx}\right|_{x_k=0}
\quad (k=2,3).$
Thus,
\begin{equation}\label{r-deriv22}
a_1^2u_{1xx}(0,t)=a_2^2u_{2xx}(0,t)=a_3^2u_{3xx}(0,t),
\end{equation}
and similarly, taking  successive $x-$derivatives of (\ref{eq1}),
we obtain the conditions on higher-order space-derivatives.
It should be emphasized that the momentum conservation
requires (\ref{r-momen3})-(\ref{r-deriv1}), from which (\ref{r-deriv22}) follows.
Now we shall prove that equations (\ref{r-cont}), (\ref{r-deriv1}) and (\ref{r-deriv22}) give a scaling function
which guarantees the infinite number of constants of motion.

Let us
introduce two functions defined on the bonds from 1 to $k (=2,3)$ as
$v_{1 \rightarrow k}\equiv u_1(\frac{a_1 x}{\sqrt{\beta}},
\frac{t}{\sqrt{\beta}})$ for $x<0$ and $\equiv u_k(\frac{a_k
  x}{\sqrt{\beta}}, \frac{t}{\sqrt{\beta}})$ for $x \ge 0$. Thanks to
the vertex boundary conditions (VBC) in
(\ref{r-deriv1}) and (\ref{r-deriv22}), together with the continuity
condition in (\ref{r-cont}), both of $v_{1 \rightarrow k}$ with
$k=2,3$ $\in C^{2}(-\infty, \infty))$ and satisfy $v_{1
  \rightarrow 2}=v_{1 \rightarrow 3}=v(x,t)$, where $v(x,t)$ is a
solution of the dimensionless sine-Gordon equation
\begin{equation}\label{sg-nf}
v_{tt}-v_{xx}+\sin v =0
\end{equation}
defined on the real line.
This fact is identical to the expression of $u_k (x,t)$ in terms of the function $v$ as
\begin{equation}\label{r-scale}
u_k(x,t)= v\left(\frac{\sqrt{\beta}}{a_k}x, \sqrt{\beta}t  \right), x \in B_k \quad (k=1,2,3).
\end{equation}

The scaling function $v$ in (\ref{r-scale}) together with
the sum rule (\ref{r-sum})
guarantees the infinite number of constants of motion and thereby the complete integrability
of the sine-Gordon equation on the network.

In fact, from \re{r-scale} and the sum rule \re{r-sum} we find that all the
conservation laws \ci{Sanuki,Geicke}
\begin{equation}\label{i1}
\int\limits_{-\infty}^{+\infty}
g(v,\pa_x v,\pa_t v, \pa_x^2 v,...,\pa_x^{n}\pa_t^{l} v)dx=const
\end{equation}
of the sine-Gordon
equation \eqref{sg-nf} on the real line also
hold on the star graph, because
\begin{align}
\sum_{k=1}^3& \int\limits_{B_k} g\left(
u_k,a_k \beta^{-\frac{1}{2}}\pa_x u_k, \beta^{-\frac{1}{2}}\pa_t u_k,...,a_k^{n}\beta^{-\frac{n+l}{2}}\pa_x^{n}\pa_t^{l}
u_k\right)dx\nonumber\\
=&a_1\int\limits_{-\infty}^{+\infty}
g\left(v,\pa_x v,\pa_t v,\ldots,\pa_x^n\pa_t^l v\right)dx\nonumber\\
&+(a_2{+}a_3{-}a_1)\int\limits_{0}^{+\infty}
g\left(v, \pa_x v, \pa_t v, \ldots, \pa_x^n\pa_t^l v\right)dx\nonumber\\
&=const.\label{i2}
\end{align}
The conservation of energy $E$, charge $Q$ and momentum
$P$ are just special cases.

From now on,
we shall prescribe $\beta=1$ without loss of generality.
Eq.(\ref{sg-nf}) has a number of explicit soliton solutions,
for instance: kink ``+'' and
anti-kink ``-'' solutions which can be written as \ci{drazin,ablowitz}
\be v(x,t)= 4
\arctan\left[ \exp\left(\pm\frac{x-x_0-\nu
      t}{\sqrt{1-\nu^2}}\right)\right], \lab{kink}
\ee where $|\nu|<1$ is the velocity of the kink. Other soliton
solutions include breathers, kink-kink collisions and
kink-antikink collisions \ci{ablowitz}, to name just a few, see
also \ci{New6} for further multi-soliton type solutions. If  the
sum rule in \eqref{r-sum} holds, then all these solutions, or,
more generally, all solutions of \eqref{sg-nf}, transfer via
\eqref{r-scale} to solutions on the metric graph. For instance,
the kinks then provide reflectionless transmission of energy
through the graph vertex, where the speed and energy of a kink
traveling in positive direction ($\nu>0$) split according to the
ratios $a_2/a_1$ and $a_3/a_1$, respectively. On the other hand,
launching two suitably fine tuned kinks on bonds 2 and 3 in
negative direction, their joint energy is transmitted to bond 1.

Before entering into the numerical analysis of kink dynamics, we comment
on other VBCs originating in the
local scattering properties at each vertex.
The VBC (\ref{r-cont}) of continuity and (\ref{VC-E}) of local flux conservation
with $\beta_k=1 (k=1,2,3)$ are also called $\delta$ VBC.
They naturally appear (\cite{caputo14}, see
also \cite{UGSBM15} for a similar construction for the case of the
NLS, and \cite[Chapter 8]{EK15} for an overview of related linear results)
by considering
the 2D sine--Gordon equation $\pa_t^2 u-\Delta u+\sin u=0$
on a ``fat'' graph, i.e., a 2D branched domain
with Neumann boundary conditions,
where $w_2/w_1=a_2^2/a_1^2$ and $w_3/w_1=a_3^2/a_1^2$ are the relative widths
of the (fat) bonds.

Similarly, the so--called $\delta^\prime$ VBCs  \cite[Chapter
8]{EK15} consist of (\ref{r-deriv1}) and \be
a_1u_1(0,t)-a_2u_2(0,t)-a_3u_3(0,t)=0, \label{dpvbc2} \ee which
conserve charge and energy for all values of the $a_k$. A simple
calculation shows  both $\delta$ and $\delta^\prime$ VBCs can be
derived from (\ref{r-sum}) and (\ref{r-scale}), but the inverse
derivation is not possible. We note that (\ref{r-deriv1}) and
(\ref{dpvbc2}) conserve $E$ and $Q$, but if conservation of $P$ is
enforced, then Eq.\re{dpvbc2} reduces to \re{r-cont}. Most
importantly, (\ref{r-sum}) and (\ref{r-scale}) give the existence
of the infinite number of constants of motion (as shown in
\eqref{i1},\eqref{i2}), which is equivalent to the complete
integrability of the sine--Gordon equation on the graph.

\textit{Vertex transmission.}
An important issue for wave dynamics in networks is the scattering
at vertices. The sum rule in \eqref{r-sum}
allows the tuning of the vertex scattering to achieve
reflectionless transmission. We now give numerical solutions of \eqref{eq1} with
$\beta_k=1 (k=1,2,3)$, using 2nd order in space and time finite differences,
 where we first focus
on \eqref{r-cont} and \eqref{VC-E} as VBC, i.e., the $\delta$ case.
Figure 2 shows the reflectionless propagation of a kink in the special
case that the sum rule \eqref{r-sum} holds.
\def\ig{\includegraphics}
\begin{figure}[ht!]
 \begin{tabular}{ll}
   (a)&(b)\\
   \ig[width=38mm,height=30mm]{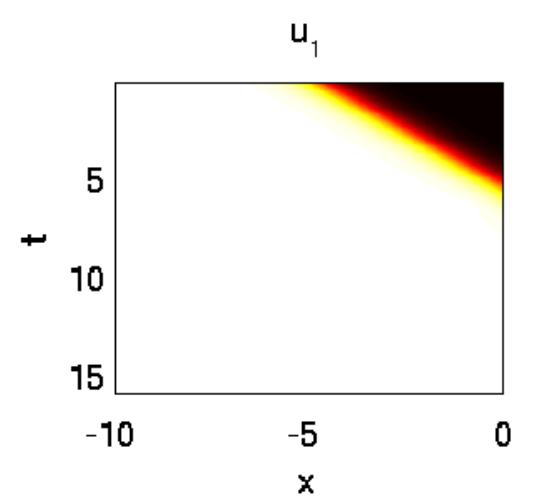}&\ig[width=51mm,height=30mm]{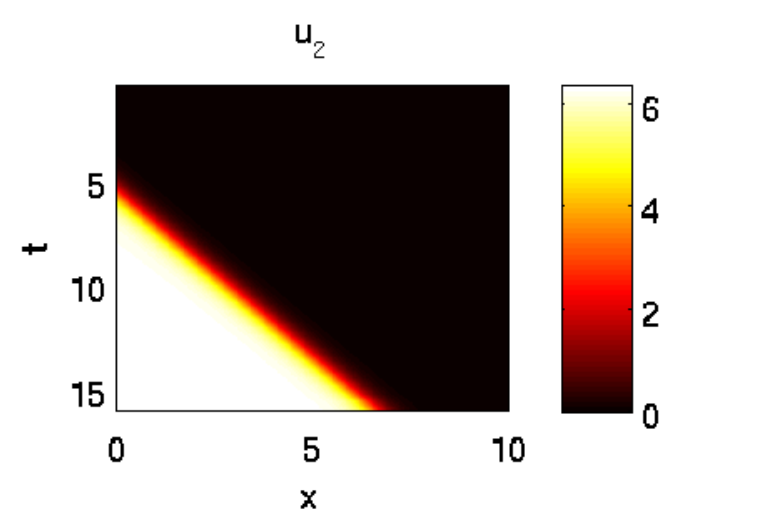} \\
   (c)&(d)\\
  \ig[width=38mm]{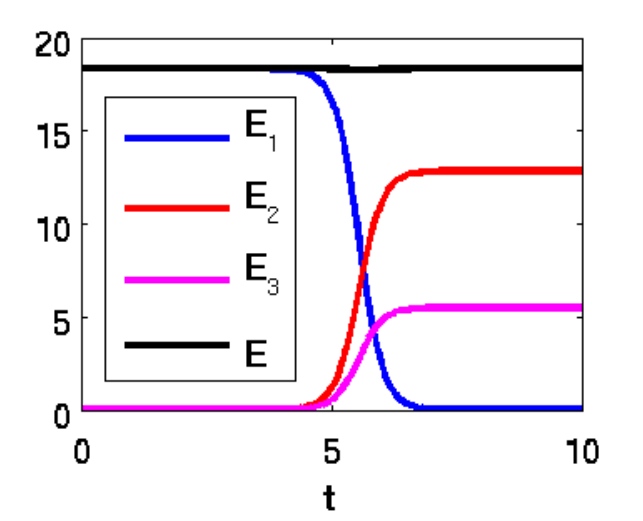}&\ig[width=38mm,height=30mm]{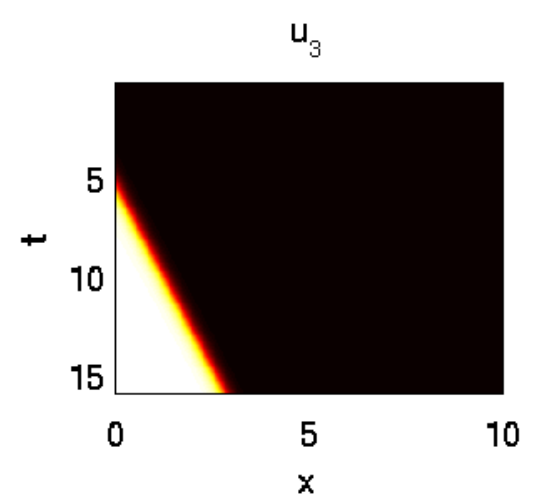}
 \end{tabular}
  \caption{{(Color online). Numerical solution of
\eqref{eq1}, \eqref{r-cont}, \eqref{VC-E},
 with  $\beta_k=1$, $k=1,2,3$,
and  $a_1=1$, $a_2=0.7$, $a_3=0.3$ fulfilling \eqref{r-sum}.
The initial conditions belongs to the kink \eqref{kink} on bond 1
 with $x_0=-5$ and
$\nu=0.9$, while $u_{2,3}=\pa_t u_{2,3}\equiv 0$.
Panels (a),(b),(d) arranged corresponding
to Fig.~\ref{pic1}, while (c) shows the $t$--dependence of the energies.
\label{pic63}}}
\end{figure}

\begin{figure}[ht!]
\begin{tabular}{ll}
   (a)&(b)\\
 \ig[height=30mm,width=38mm]{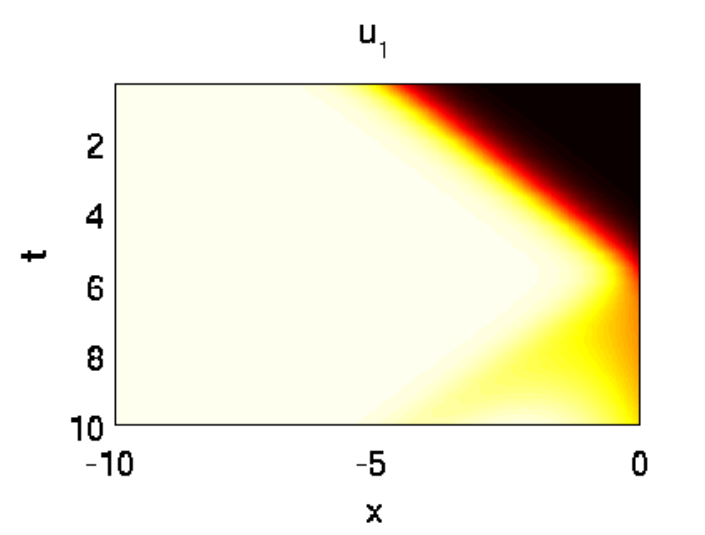}&\ig[height=30mm, width=48mm]{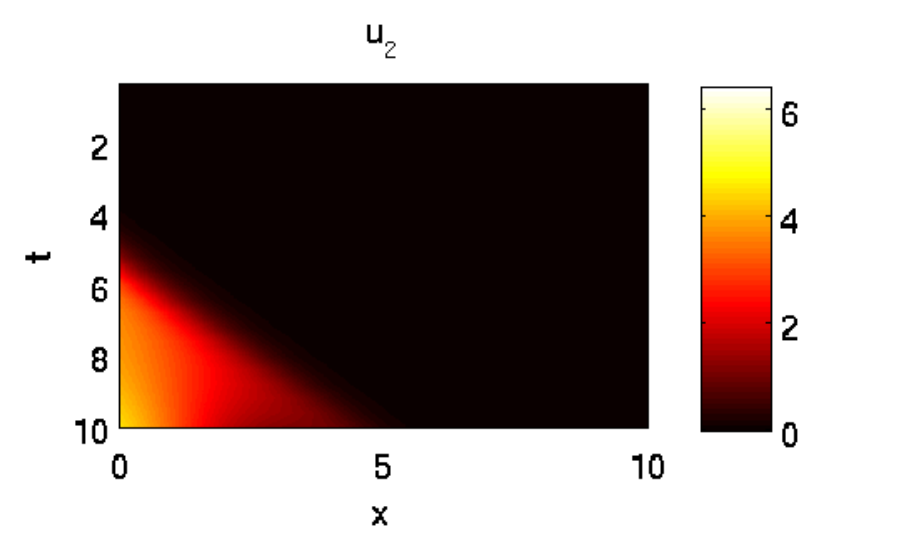}\\
(c)&(d)\\
 \ig[height=30mm,width=38mm]{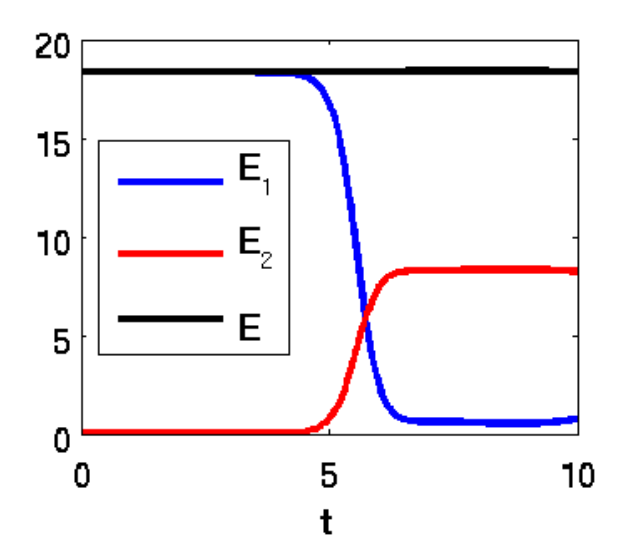}&\ig[height=30mm, width=42mm]{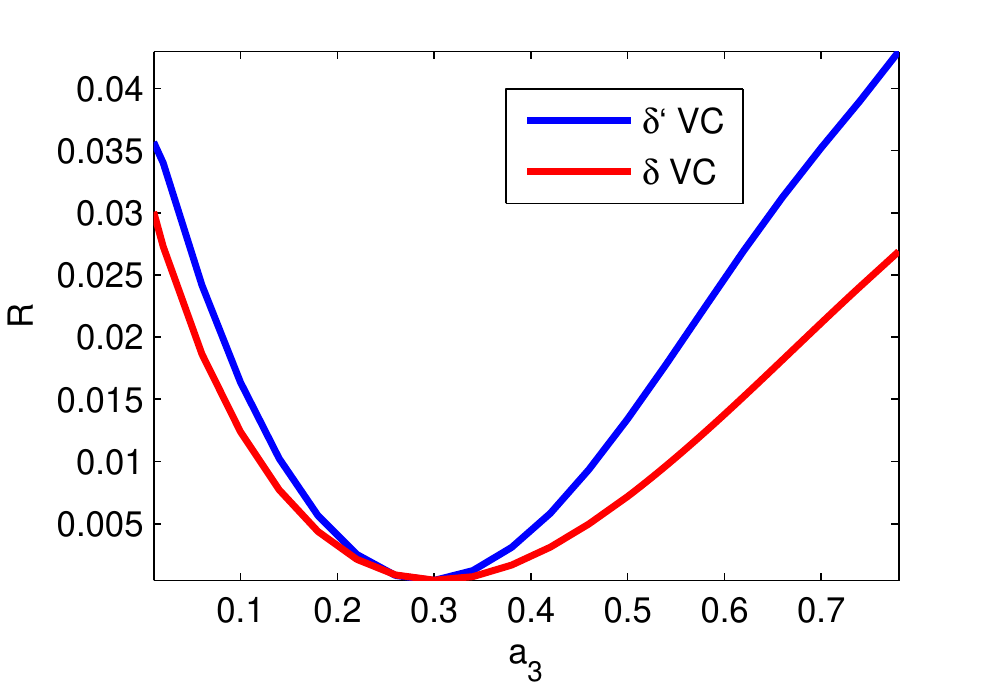}
\end{tabular}
  \caption{{(Color online). (a)-(c)  Numerical solution of
\eqref{eq1},  \eqref{r-cont}, \eqref{VC-E}, with $\beta_k=1$ for $k=1,2,3$.
(a)-(c) Reflection of incoming kink at the
vertex in case that $a_1=a_2=a_3=1$, violating \eqref{r-sum},
initial conditions as in Fig.~\ref{pic63}.
$u_3$ is identical to $u_2$, and hence $u_3$ and $E_3$ in (c) are omitted.
(d) (blue line) Dependence of the  vertex reflection coefficient
$R=E_1|_{t=15}/E_1|_{t=0}$
on $a_3$, where $a_1=1, a_2=0.7$,
hence $a_3=0.3$ corresponding to Fig.~\ref{pic63}. The red line is similarly
obtained from solving \eqref{eq1} with
$\delta'$ VBC \eqref{r-deriv1} and \eqref{dpvbc2}.
\label{pic66}}}
\end{figure}

In Fig.~\ref{pic66} we numerically treat the transmission
of solitons through the graph vertex when the sum rule (\ref{r-sum})
is violated. 
In (a)-(c) we consider the ``natural'' case
$a_k{=}1$, $k=1,2,3$, and  the same kink initial condition as in Fig.\ref{pic63}.
 The total energy is still conserved (by \eqref{VC-E}), but
in contrast to the reflectionless case from Fig.\ref{pic63}, there now is
significant reflection at the vertex.
To demonstrate and quantify the dependence of the vertex transmission on the
$a_k$ in some more detail, in Fig.~\ref{pic66}(d) we essentially return
to the situation of Fig.~\ref{pic63}. That is, we set $a_1=1$, $a_2=0.7$,
but let
$a_3$ vary and plot the reflection coefficient $R$, defined as the
ratio of the energies in the incoming bond at initial time $t=0$ and
at $t=15$. At $a_3=0.3$, corresponding to Fig.~\ref{pic63}, we have $R=0$,
i.e.~zero reflection.

\begin{figure}[h!]
\centerline{\includegraphics[width=100mm]{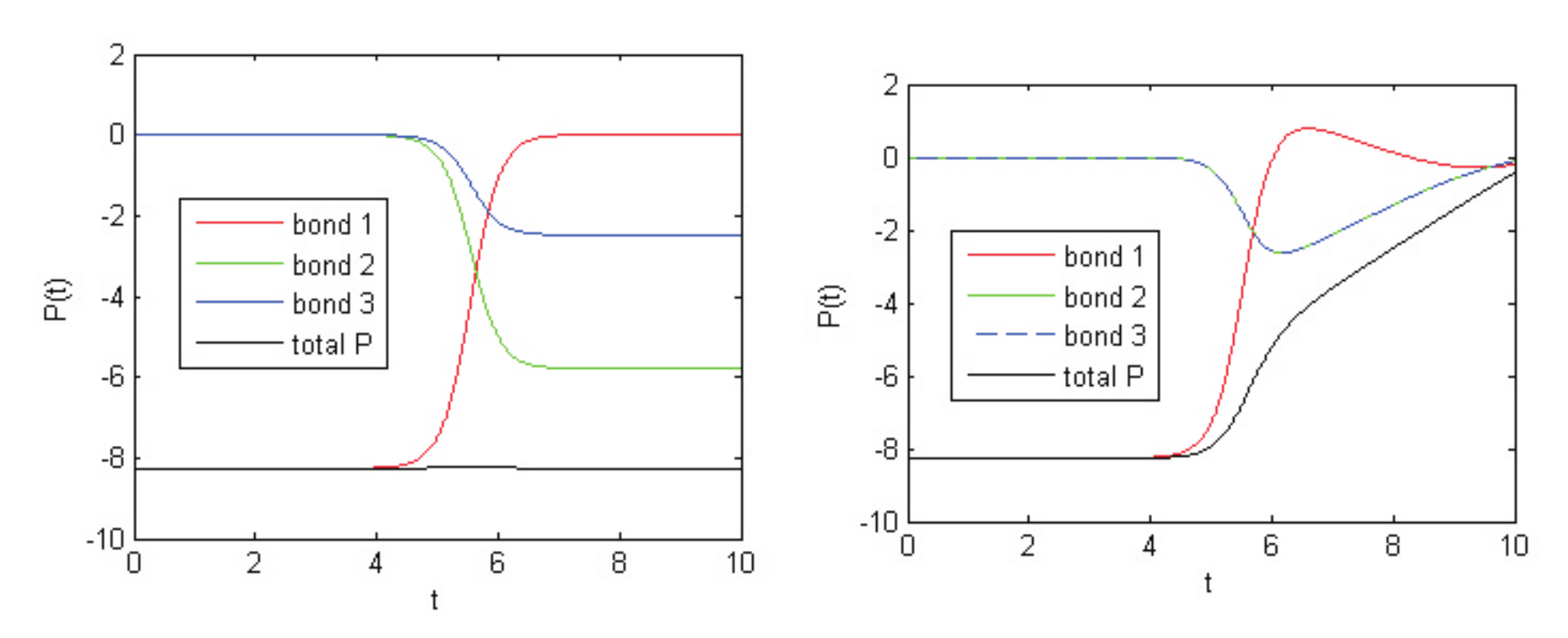}}
\caption{(Color online). Time evolution of momenta. Left
panel: $a_1 = 1,~ a_2 = 0.7,~ a_3 = 0.3$; right panel: $a_1 = a_2
= a_3 = 1$.} \label{pic5}
\end{figure}

Additionally, the red line in Fig.~\ref{pic66}(d) shows the
analogous simulation for the case of $\delta'$ vertex conditions
\eqref{r-deriv1} and \eqref{dpvbc2}. Again we have zero reflection
at $a_3=0.3$, while violating the sum rule gives qualitatively
similar but slightly stronger reflections than the $\delta$ case.
Two points should be noted: the simulations in Figs.~2 and 3 have
also confirmed the conservation up to numerical
discretization effects) of $Q$ and $P$ so long as the sum rule
(\ref{r-sum}) holds (see Fig.\ref{pic5}); the
simulations do not use the soliton properties of the kinks, but
only the fact that they are traveling wave solutions for which we
have formulas for the initial conditions. Thus, these numerical
results  can be transfered to general nonlinear Klein-Gordon
 equations that admit travelling wave solutions.

\begin{figure}[htb]
  \centerline{\includegraphics[width=60mm]{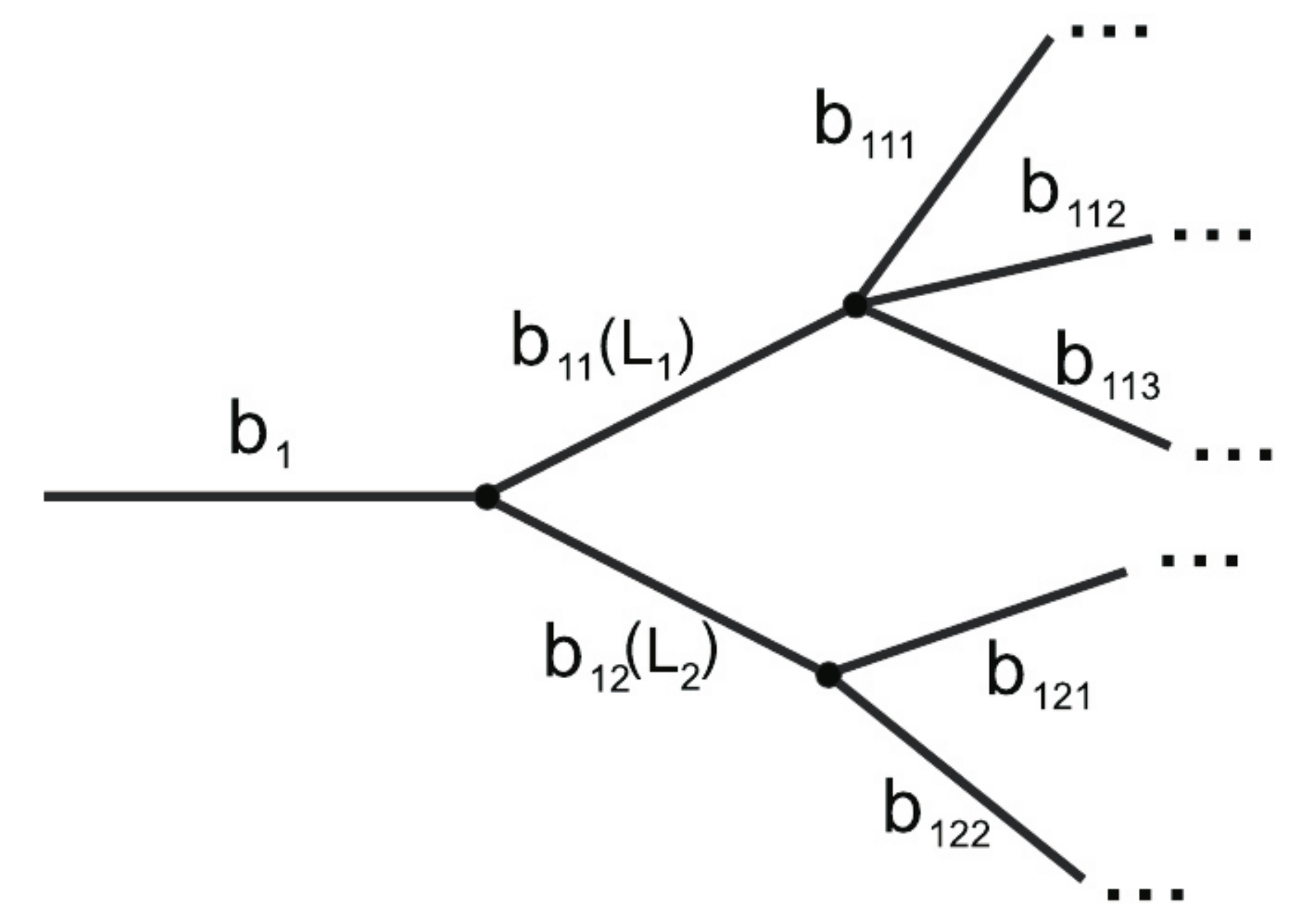}}
\caption{A tree
    graph with three layers, $b_1{\sim}(-\infty, 0), b_{11}, b_{12} \sim (0,L_k)$, $k=1,2$, and
    $b_{1ij} \sim (0,+\infty)$ with $i,j=1,2, \ldots$.} \label{tree-0}
\end{figure}

\textit{Other graph topologies.} Our results can be extended to
other simple topologies such as general star graphs, tree graphs,
loop graphs and their combinations.
 Exact traveling wave solutions
of sine-Gordon models on such graphs with one incoming
semi-infinite bond can be obtained similarly to the above case of
a star graph with three bonds, leading to generalizations of the
sum rule. We illustrate this for the tree graph from
Fig.~\ref{tree-0}, consisting of three ``layers'' $b_1, (b_{1i}),
(b_{1ij})$, where $i,j$ run over the given bonds.

On each bond $b_1, b_{1i}, b_{1ij}$ we have a sine-Gordon equation
given by (\ref{eq1}). Setting $\beta_1=\beta_{1i}=\beta_{1ij}=1$
for all $i,j$, the $a_{1i}$ and $a_{1ij}$  have to be determined
from the sum rule like \eqref{r-sum} at each vertex.
For instance, at the three nodes in Fig.~\ref{tree-0} we need
\begin{equation}\label{r0}
\begin{array}{ll} \text{ end of } b_1:\quad&a_0=a_{11}+a_{12},\\
\text{ end of } b_{11}:\quad&a_{11}=a_{111}+a_{112}+a_{113},\\
\text{ end of } b_{12}:\quad&a_{12}=a_{121}+a_{122},
\end{array}
\end{equation}
and this continues through the layers. By (\ref{sg-nf}) and (\ref{r-scale})
this is based on scalings such as
\begin{equation}\label{r1}
u_1(x,t)=v(x/a_1,t) \text{ and } u_{1i}(x,t)=v(x/a_{1i},t),
\end{equation}
where at subsequent bonds we also need to take into account
the finite propagation length in the previous bonds, for instance
\begin{equation}\label{r2}
u_{111}(x,t)=v((x+x_0)/a_{111},t),\
x_0/a_{111}=L_1/a_{11},
\end{equation}
i.e.~$x_0=a_{111}L_1/a_{11}$. Necessarily, the speeds and
energies of, e.g., an incoming kink,
also split according to rules like \eqref{r0}, such that
on each final bond we only have slow and small energy kinks.
A similar construction has been done and formalized for
the propagation of Nonlinear Schr\"odinger solitons on
tree graphs in \cite{zar2010}.

\begin{figure}[htb]
 \centerline{\includegraphics[width=\columnwidth]{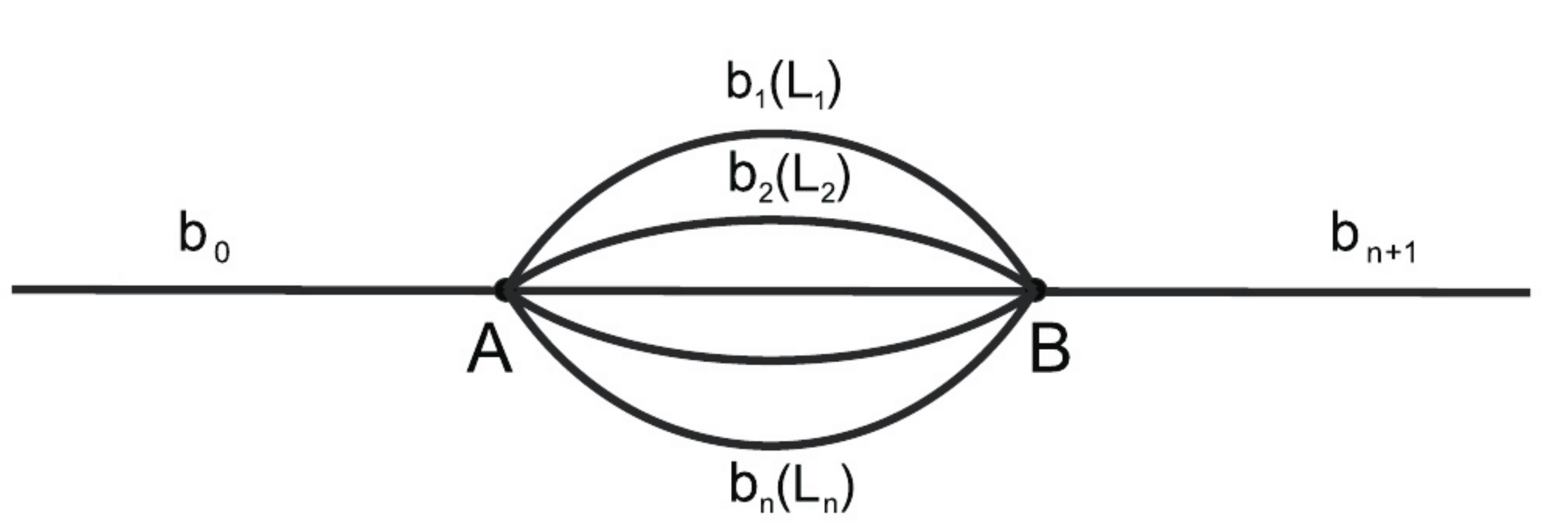}}
  \caption{A graph with a loop. $b_0\sim (-\infty, 0),
 b_k \sim (0,L_k)$, $k=1,\ldots,n$, where $L_k=a_kL$,
$b_{n+1}\sim (0,\infty)$.} \label{loop}
\end{figure}
Another graph for which soliton solutions of sine-Gordon models
can be obtained is a graph with a loop (see Fig.\ref{loop}), which
consists of two semi-infinite bonds connected by $n$
bonds having finite lengths $L_k$. Requiring the conditions
\begin{equation}\label{r4}
a_0=\sum_{k=1}^{n}a_k=a_{n+1}
\end{equation}
for the coefficients,
and $L_k=a_k L$ ($k=1,2,...n$)  with a constant $L$,
we can write soliton solutions in a similar way as in \eqref{r1} and \eqref{r2}.

Finally, it can be shown that the above approach can be applied to
obtain exact traveling wave solutions of sine-Gordon models on
other (than above) graphs consisting of at least two semi-infinite bonds and any
subgraph between them.  In this case one needs to impose the pertinent
vertex conditions like \eqref{r1} or \eqref{r4} at the vertices connecting the
semi-infinite bonds with the subgraph.  \medskip

\textit{Conclusions.}
In this work we studied sine-Gordon equations on simple metric
graphs, and derived vertex boundary conditions for charge, energy
and momentum conservation, and additionally conditions on parameters,
for which the problem has explicit analytical soliton solutions.
We find the sum rule \re{r-sum} for bond-dependent coefficients at each
vertex of the graph, which makes the sine-Gordon equation on the graph
 completely integrable.
It is shown that the obtained solutions provide the reflectionless transmission
of solitons at the graph vertex.
 This is also illustrated numerically by
quantifying the reflections for a case where these conditions are
violated, and we discussed how to generalize the results to other graph topologies.
The results can be
directly applied to several important problems such as Josephson junction network and
DNA double helix.
 In such  approach our model corresponds to continuous version of the system considered in \ci{sodano-np09,sodano13}.
Finally, a very important application can be DNA double helix models where the
energy transport is described in terms of sine-Gordon equations
\ci{Yamosa,Yakushevich1}. Base pairs of the DNA double
helix can be considered as a branched system and modeled by a
star graph \ci{Yakushevich1}. Then the $H$-bond energy between two base pairs in such system can be characterized by the parameter, $\beta$.

{\it Acknowledgement.}  We thank Panayotis Kevrekidis for his valuable comments on this paper. This work is supported by a grant of the
Volkswagen Foundation. The work of DM is partially supported by the grant of the Committee for the Coordination Science and Technology Development (Ref.Nr. F3-003).

\bibliography{biblar}
\bibliographystyle{apsrev4-1}

\end{document}